\documentclass[preprint,12pt]{elsarticle}
\usepackage[english]{babel}
\usepackage[T1]{fontenc}
\usepackage[latin9]{inputenc}
\usepackage{float}
\usepackage{amsmath}
\usepackage{amssymb}
\usepackage{graphicx}
\usepackage{verbatim}
\usepackage{algorithm}
\usepackage{algorithmic}
\usepackage{times}
\usepackage{epstopdf}
\usepackage{amssymb}
\usepackage{amsmath}
\usepackage{stfloats}
\usepackage{color}
\usepackage{multirow}
\usepackage{array}
\newcommand{\PreserveBackslash}[1]{\let\temp=\\#1\let\\=\temp}
\newcolumntype{C}[1]{>{\PreserveBackslash\centering}p{#1}}
\newcolumntype{R}[1]{>{\PreserveBackslash\raggedleft}p{#1}}
\newcolumntype{L}[1]{>{\PreserveBackslash\raggedright}p{#1}}

\def\papervec{\tt Paper2Vec}

\makeatletter

\floatstyle{ruled}
\newfloat{algorithm}{tbp}{loa}
\providecommand{\algorithmname}{Algorithm}
\floatname{algorithm}{\protect\algorithmname}

\@ifundefined{showcaptionsetup}{}{%
 \PassOptionsToPackage{caption=false}{subfig}}
\usepackage{subfig}
\makeatother

\journal{Neurocomputing}
\begin{document}

\begin{frontmatter}
\title{Paper2vec: Citation-Context Based Document Distributed Representation for Scholar Recommendation}

\author{Han Tian}
\ead{tianhan\_4@aliyun.com}
\author{Hankz Hankui Zhuo\corref{cor}}
\ead{zhuohank@gmail.com}

\address{School of Data Science and Computer, Sun Yat-sen University, Guangzhou, China}

\cortext[cor]{Corresponding Author\corref{cor}}

\begin{abstract}
Due to the availability of references of research papers and the rich
information contained in papers, various citation analysis approaches
have been proposed to identify similar documents for scholar recommendation.
Despite of the success of previous approaches, they are, however,
based on co-occurrence of items. Once there are no co-occurrence items
available in documents, they will not work well. Inspired by distributed
representations of words in the literature of natural language processing,
we propose a novel approach to measuring the similarity of papers
based on distributed representations learned from the citation context
of papers. We view the set of papers as the vocabulary, define the
weighted citation context of papers, and convert it to weight matrix
similar to the word-word co-occurrence matrix in natural language
processing. After that we explore a variant of matrix factorization approach to train distributed representations
of papers on the matrix, and leverage the distributed representations to measure similarities of papers. In the experiment, we exhibit that our approach outperforms state-of-the-art citation-based approaches by 25\%, and better than other distributed representation based methods. 
\end{abstract}

\end{frontmatter}

\section{Introduction}

\begin{table}[tbp]
\small
\begin{tabular}{|c|c|c|c|c|c|c|}
\hline
 & WoS & Scopus & CiteSeerX & DBLP & PMC & arXiv \\
\hline
Full text availability & No & No & Yes & Yes & Yes & Yes \\
\hline
Records in millions & $\sim90$ & $\sim55$ & $\sim6$ & $\sim3$ & $\sim3$ & $\sim1$ \\
\hline
\end{tabular}
\caption{\label{tab:datasets}List of some popular datasets. Citation index often contains much more records than full-text dataset.}
\end{table}

Recommender systems have been introduced into many academic services,
such as CiteSeerX%
\footnote{http://citeseerx.ist.psu.edu/%
}, Google Scholar%
\footnote{http://scholar.google.com%
}, PubMed%
\footnote{http://www.ncbi.nlm.nih.gov/pubmed%
}, and scholar social network such as ResearchGate%
\footnote{https://www.researchgate.net%
}, reference managers such as CiteULike%
\footnote{http://www.citeulike.org%
}, Docear%
\footnote{http://www.docear.org/%
}, Mendeley%
\footnote{https://www.mendeley.com/%
}. Due to the availability of paper references, many approaches based
on citation analysis have been proposed to enhance the performance
of relevant-document search \cite{meuschke2015citrec}. Reseachers found document retrieval methods using citation linkages are able to find additional relevant documents than conventional word indexing methods \cite{leepao1993term}. While full text documents are not always open access, citation indexes such as Web of Science, Google Scholar and Microsoft Academic Search can track citation linkage for most papers. Table \ref{tab:datasets} demonstrates some popular scholar datasets and citation indexes.

Most of citation based methods view the number of co-occurrence of the citation linkages as similarity measurement via considering different citation linkage types with different weighting schemes. In those approaches, they require that there is at least one item shared in the contexts of two papers in order to calculate their similarity \cite{Beel2015}.
However, it is common for lots of pairs of documents that are similar but having no shared citation linkages. It may be caused by the fact that they come from different sources: technical reports, books, case-report and so on, or the time span between two papers are too long or too short. Table \ref{tab:egPapers} demonstrates some examples from dataset.

In this paper, we present a novel approach, called {\papervec},
indicating that each paper is represented by a real valued vector.
Inspired by distributed representations of words proposed in the area of
NLP, which have recently demonstrated state-of-the-art results across
various NLP tasks, we view each scholar paper (or paper ID specifically) as a word, and learn the distributed representations of words based on the citation context of papers, to capture the implicit scholar topics contained
in the citation linkage set. Our paper distributed vectors are trained in a stochastic
way based on matrix factorization on the citation relations data. And the cosine similarity of vectors is used as the document similarity measurement to find relevant scholar
papers. The stochastic training way also makes Paper2vec easy for online learning. 

\begin{table}[tbp]
\centering
\small
\begin{tabular}{C{6cm}|C{6cm}}
\hline
\multirow{2}{*}{\begin{tabular}{c}\emph{Nonmelanoma skin cancer in}\\\emph{India: Current Scenario}\end{tabular}} & \multirow{2}{*}{\begin{tabular}{c}\emph{Perianal Basal cell carcinoma-an} \\\emph{unusual site of occurrence}\end{tabular}} \\
& \\
\hline
\multirow{3}{*}{\begin{tabular}{c}\emph{Expression profiles during} \\\emph{honeybee caste determination}\end{tabular}} & \multirow{3}{*}{\begin{tabular}{c}\emph{Semiparametric approach to} \\\emph{characterize unique gene expression} \\\emph{trajectories across time}\end{tabular}} \\
& \\
& \\
\hline
\multirow{2}{*}{\begin{tabular}{c}\emph{Ten years of general practice}\\\emph{midwifery 1954/1963}\end{tabular}} & \multirow{2}{*}{\begin{tabular}{c}\emph{Two Hundred Years of} \\\emph{Midwifery 1806 - 2006}\end{tabular}} \\
& \\
\hline
\multirow{3}{*}{\begin{tabular}{c}\emph{Accessing and distributing} \\\emph{EMBL data using CORBA}\end{tabular}} & \multirow{2}{*}{\begin{tabular}{c}\emph{integrOmics: an R package} \\\emph{to unravel relationships} \\\emph{between two omics datasets}\end{tabular}} \\
& \\
& \\
\hline
\end{tabular}
\caption{\label{tab:egPapers}Examples of pairs of papers similar but having no shared citation papers and not cited by the same paper. We find them from the PubMed Central dataset by our algorithm. For row 1, basal cell carcinoma is a kind of nonmelanoma skin cancer, while two papers are considering about different aspects. For row 2, while two papers are both about gene expression research, the former studies the bees and the latter studies the tools. Pairs in row 3 are both midwifery practice reports, but between a large span of time. Pairs in row 4 talks about different tools for similar goals.}
\end{table}

As far as we know, there is no related research based on distributed representation
for citation based algorithm. \cite{le2014distributed} also proposed a way to train documents as vectors under the framework of recent distributed representation models of NLP, however, it's based on the full text corpus of papers. 
In summary, our contributions are shown as follows.
\begin{itemize}
\item we can calculate the similarity between any pair of document
without the need of intersection of citation linkage sets. 
\item
full text is not needed for Paper2vec, which makes it possible to
be applied into scholar databases where full text is not supported.
\item the stochastic learning process and the corpus structure make
it possible an online learning process. When a new paper is included
into the database, it can be transformed into training data and learned
immediately.
\end{itemize}

The paper is organized as follows. In Section 2 we review the related
work of Paper2vec, from the citation-based similarity measures (Section
\ref{sub:Citation-based-algorithms}) to the word distributed representation
training algorithms (Section \ref{sub:Word2vec}). Section 3 describes
Paper2vec in details and the similarity between paper vectors and
word vectors. Section 4 contains the details of the evaluation experiment,
and Section 5 draws some conclusions and addresses future aspects
of our work.

\section{Related Work}

\subsection{\label{sub:Citation-based-algorithms}Citation-based Algorithms}

Many different similarity measures were proposed derived from document
citation structure. A lot of research have proved that the search
performance can be enhanced by incorporating citation algorithms into
IR systems \cite{eto2013evaluations}. Among them the most widely
used three basic methods are Co-citation, Bibliographic Coupling and
Amsler, invented in the 60s and 70s \cite{cristo2003link}. They calculate
the intersection of different citation linkage sets. While Co-citation
regards the times two documents are cited together, namely ``co-cited'',
as the similarity measure, Bibliographic Coupling consider the number
of documents they share in their references. To combine the two basic
algorithms to get better results, Amsler proposed an algorithm considering
the intersection of the union of two citation linkage sets mentioned above. Pairs
of documents under all three models cannot be compared without co-occurrence
items.

Context information of citation were introduced recently into the
co-citation based similarity measure with different weighting schemes
to quantify the degree of relevance between co-cited documents. Citation
Proximity Analysis (CPA) \cite{gipp2009citation}, for instance, takes
fixed value reflecting the proximity between two citations in the
full text as the strength of the relevance of two co-cited papers,
while \cite{callahan2010contextual} use another proximity function
to get co-citation strength based on document structure. However,
they have the same problems as the classical methods do, and the need
of full text of such context-based co-citation methods limits their
availability in some large datasets such as Web of Science, where
full text is not supported.

\subsection{\label{sub:Word2vec}Word Distributed Representation Models}

The distributed representation for words were initially applied to
solve the curse of dimensionality caused in language modeling based
on discrete representation of words \cite{bengio2003neural}. Under
the hypothesis that words having similar context are similar, which
suggests that contextual information nicely approximates word meaning,
distributional semantic models (DSMs) use continuous short vectors
to keep track of the context information collected in large corpus,
namely the word distribution nearby the specific word. \cite{baroni2014don}
classified DSMs into two types, count models and predict models. While
count models are the traditional methods to generate distributed vectors
by transforming the feature vectors with various specific criteria,
predict models build a probabilistic model based on word vectors,
which are trained on large corpus to maximize the probability of the
occurrences of the contexts of observed words. Because the probability
function is smooth and calculated totally by the distributed word
vectors instead of discrete word features, the word distributed representation
obtained by predict models makes it possible to regard two words similar
when the distribution of context word vectors are similar. Evaluation
performed in \cite{baroni2014don} shows context-predicting models
are superior than count-based models on several measures.

Among existing predict models, SkipGram uses a simple probability
function and achieved promising performance. SkipGram was proposed
in \cite{Mikolov2013c} and improved in \cite{mikolov2013distributed},
which tries to predict the context of observed words $P(context\vert w)$
across the corpus. The context of word is defined as a window around
the word, the window size can be parameterized. Some efficient approximation
methods are proposed to accelerate the training process such as hierarchical
softmax and negative sampling \cite{Mikolov2013c}. The training process
is stochastic and iterates several times across the large corpus.
\cite{NIPS2014_5477} proved SkipGram with negative sampling is implicitly
factorizing a shifted point-wise mutual information (PMI) transformed
word-context occurrence matrix. \cite{Mikolov2013c} have also proposed
another algorithm named CBOW, which is an approximation version of
SkipGram. 

Different from SkipGram, GloVe \cite{pennington2014glove} minimize
the cost function corresponding to the probabilistic function SkipGram
maximizes, which is based on the word-word co-occurrence matrix collected
from training corpus. The nonzero elements on the matrix are trained
to obtain word representations. This model efficiently leverages
global statistical information and connects local context window methods
such as SkipGram to the well-established matrix factorization methods, our algorithm is inspired by it.

\section{\label{sec:Paper2vec}Paper2vec}

The latent vertex dimensions learned should be continuous and informative
of the context. In this section we describe the Paper2vec algorithm,
which learns distributed vertex embeddings from matrix factorization
on the weighted context definition of node. 

\subsection{Weighted Citation Link Context}

\begin{algorithm}[tbp]
\caption{Framework of Paper2vec algorithm}
\label{alg:algorithm1} 
\begin{algorithmic}[1] 
\REQUIRE The scholar database of research papers $D$ containing
citation relation; 
\ENSURE Distributed embeddings of research papers contained in $D$,
$W$; 
\STATE Build the citation relation network from $D$;
\STATE Construct the citation linkage weight matrix of research
papers from the network; 
\STATE Minimize the cost function stochastically to get the paper vectors $W$;
\RETURN $W$; 
\end{algorithmic} 
\end{algorithm}

Many citation based similarity measurements are based on the assumption that papers having similar citation relation are similar. If we model the citation relation of papers to a directed graph, where nodes representing the documents, links representing citation relations, a classical similarity measurement approach is to measure the intersection of sets of neighbor nodes of the compared nodes representing the compared documents \cite{cristo2003link}. The set implicitly demonstrates the semantic content of the target paper. We define the set as ``citation link context'' of a document because the context set is based on the citation relation of papers%
\footnote{The concept ``citation context'' has been used to describe the text
surrounding the citation position in full text in previous research. %
}. 

However, we find that the citation link context is not limited to the direct citation relation, but can extend to papers having indirectly citation relation with the target paper, which also implicit the semantic context, with a weaker weight. Our first task is to define a new weighted citation link context that extends the neighbor nodes, which can help us better measure the similarity between documents. 

The weight scheme of citation link context should consider the following characteristics we observe in the scholar dataset:
\begin{itemize}
\item Cited papers and citing papers together help predicting the content
of the current paper. While cited papers reflect the topics at the
publication time, citing papers focus more on the academic significance
afterwards. 
\item Decrease with distance. Papers indirectly cited are not so relevant
than the directly cited papers and have smaller weight.
\item Transitivity. Weights can be transmitted. Assume paper A cites paper B and paper B cites paper C, weaker the weight of citation link between A and B or B and C, weaker the weight of link between A and C.
\end{itemize}

To satisfy the above properties, we define a weight scheme based on random walk propability. We don't treat cited link and the citing link differently. So we can get a undirected graph from the citation relation of dataset. 
In the graph, we consider the probability randomly walking from node $A$ to node $B$ as the weight of B to A, which implicit the weight scheme is not symmetric. 
To simplify the calculation, we only concern nodes in a predefined window $win$, meaning only taking the nodes having less than $win$ steps away from the target node along the link path into account. Farther nodes may have too small to have an impact for the result.
We define $A$ as the transition matrix of nodes, $A_{ij}$ meaning the probability walking from node $i$ to node $j$. The transition matrix only consider the probability between neighbor nodes, and the probability from node to its neighbors are equally allocated. Then the weight
can be calculated as follows:

\begin{equation}
X(j \mid i)=max(0,\log{[\sum_{o=1}^{win}\sum_{k=1}^{o}A^{k}]_{ij}}+\lambda)
\end{equation}

The weight is the shifted positive logarithm of expected time we arrive node $j$ when random walk
from node $i$ for $win$ steps. $X(j \mid i)$ represents the weight of node $j$ for node $i$. We use logarithm function to change the exponential decay of weight with respect to the distance to the target node to linear decay, then we shift it to get positive weights, which are asymmetric and not fixed. The parameter $\lambda$ should 
be chose based on the dataset and the window to guarantee most of the weight information are reserved. 
With the weight scheme we define a new richer citation link context, which can help us find a better similarity measurement.

\subsection{\label{sub:Learning-Vertex-Dimensions}Learning Vertex Dimensions
with Citation Link Context}

\begin{algorithm}
\caption{Paper2vec($i$, $j$, $X(i \mid j)$)}
\label{alg:algorithm1-1} \begin{algorithmic}[1]
\STATE $grad=2*f(X(i \mid j)(w_{i}^{T}\widetilde{w_{j}}+b_{i}+\widetilde{b_{j}}-X(i \mid j))$
\STATE $temp_{i}=alpha*grad*\widetilde{w_{j}}$ ; $temp_{j}=alpha*grad*w_{i}$
\STATE $\widetilde{w_{j}}=\widetilde{w_{j}}+temp_{j}$; $w_{i}=w_{i}+temp_{i}$
\end{algorithmic} 
\end{algorithm}

Distributed representation have recently demonstrated state-of-the-art results across
various NLP tasks. The successful application is based on the assumption that words having similar context are similar. The similarity between the 
assumptions inspired us to use distributed representation to represent papers, which is learned from the weighted citation link context we get in the 
last section. While word distributed representations can implicit the semantic and syntatic information of word, we expect scholar document to be represented
by distributed vectors to capture the implicit scholar topics contained in the citation link set.

Now the question is how to utilize the weighted citation link set. We can transfer it to a sparse weight matrix $W$, 
where $W_{ij}$ representing the weight node $j$ for node $i$. Then matrix factorization approach can be used to obtain vectors representing nodes.
The cost function for training is defined as follows:

\begin{equation}
J=\sum_{i,=1}^{V}\sum_{j=1}^{V}f(X(j \mid i))(w_{i}^{T}\widetilde{w_{j}}+b_{i}+\widetilde{b_{j}}-X(j \mid i))^{2}
\end{equation}

where $w$ and $\widetilde{w}$ are respectively
the paper embeddings and the context embeddings. $f(X_{ij})$ controls the weight of elements in the matrix. While small $X(j \mid i)$ may mean less information and more noise, we give a weighting 
function $f(X(i \mid j))$ for every element in the matrix as follows:

\begin{equation}
f(X(j \mid i))=[\sum_{o=1}^{win}\sum_{k=1}^{o}A^{k}]_{ij}
\end{equation}

Notably, we use bias $b$ and $\widetilde{b}$ to loose the constraint of $\lambda$ in the weight scheme, making cost function more flexible. When the cost function is satisfied and 
according to the weight scheme, the exponential of the inner product of the paper vector and the context vector represents the random walk probability of the context paper:

\begin{equation}
\exp^{w_{i}^{T}\widetilde{w_{j}}}=\frac{[\sum_{o=1}^{win}\sum_{k=1}^{w}A^{k}]_{ij}}{\exp^{b+\widetilde{b}-\lambda}}
\end{equation} 

When the window of random walk is set to $win$, the sum of probability of all context nodes is $win$. So the exponential value of inner product represents the probability ratio. 
The cost function approximates the document embeddings from a different way compared with DeepWalk \cite{perozzi2014deepwalk}.

Given the definition of context and the weight scheme, we can minimize
the cost function to get representations representing the items.
We note that the weight is not symmetric, $X(j \mid i)\text{\ensuremath{\not\not\not}=}X(i \mid j)$,
different from the word-word co-occurrence matrix in the area of NLP, and
is often the case for the relationship between items and friends. We tried to average the weight for $X(j \mid i)$ and
$X(i \mid j)$, but the former won. The updating procedure is
described in detailed in Algorithm \ref{alg:algorithm1-1}.

The complete algorithm of Paper2vec is described in Algorithm \ref{alg:algorithm1}.
The same as GloVe, Paper2vec will use a stochastic learning way to
iterate the nonzero items in the matrix. Because the citation corpus
is not as redundant as text corpus, the iteration time is often larger
than that used in NLP tasks. After training, the context vectors are
dropped and we obtain the paper vectors after normalizing, which can
be used to measure similarity by calculating the cosine similarity
of vectors of two documents as similarity measurement:

\begin{equation}
paper2vec(d_{i},d_{j})=p{}_{i}^{T}p_{j}
\end{equation}

\section{Experiment}
Conducting a nice evaluation experiment is challenging in research-papers
recommender system, relating to the lack of datasets and gold standards
\cite{Beel2015,meuschke2015citrec}. Our experiment is conducted based on CITREC,
an open evaluation framework for citation-based similarity
measures proposed in \cite{meuschke2015citrec}, which provides scholar datasets, baseline measurement and some implementations of previous citation-based algorithms. 

\subsection{Dataset}
CITREC has collected the data from the PubMed Central Open Access Subset (PMCOS) and the TREC Genomics collection. PubMed Central is a repository of full text documents from biomedicine and the life sciences maintained by the U.S. National Library of Medicine (NLM). The NLM offers a subset of 860,000 documents for downloading and processing. TREC Genomics Collection is a test collection used in the Genomics track of the TREC conference 2006, which comprises approx. 160 thousands Open Access biomedical full text articles.

We extracted the citation relation from the full text with the methods CITREC provided and constructed a database with documents, references for both datasets. With reference information collected in full text, we conducted entity resolution between documents and references based on PubMed ids, titles and authors, et al. We collected 252673 documents and 9379146 references for PMCOS, 160446 documents and 6312425 references for TREC Genomics. In order to make datasets self-containment, we only construct distributed vectors for
papers contained in recorded documents. For other baseline methods, we also limit the available references data to the subset of that included in the recorded documents. 

\subsection{Gold Standards}
The standard similarity score is calculated based on the Medical Subject Headings thesaurus (MeSH), which are a poly-hierarchical thesaurus of subject descriptors, maintained by experts at the U.S. National Library of Medicine (NLM), and available for two scholar datasets mentioned above. CITREC include a gold standard suggested by \cite{lin1998an} based on MeSH to reflect topical relevance between documents. The similarity measurement demonstrates the proximity of the subject descriptors of two papers across the concept hierarchical tree, which can be considered as a suitable way to measure the semantic similarity between papers \cite{meuschke2015citrec}.

\subsection{Baseline Methods}
We compare our proposed method to some representative methods for
citation-based analysis and network representation learning approaches which can be transfered to this area.
\begin{itemize}
\item Amsler \cite{cristo2003link}: This model calculate the intersection
of papers having citing or cited relation with the measured pair of
papers. The similarity score can be formalized as follows:
\end{itemize}
\begin{equation}
amsler(d_{i},d_{j})=\frac{(P_{d_{i}}\cup C_{d_{i}})\cap(P_{d_{j}}\cup C_{d_{j}})}{\vert(P_{d_{i}}\cup C_{d_{i}})\cup(P_{d_{j}}\cup C_{d_{j}})\vert}
\end{equation}

$P_{d_{i}}$ in the Equation is defined as the paper set citing $d_{i}$
and $C_{d_{i}}$ the cited paper set of $d_{i}$.
\begin{itemize}
\item CPA \cite{gipp2009citation}: Context information of citation were
introduced into the this model to build co-citation based similarity
measure with different weighting schemes. To quantify the degree of
relevance between co-cited documents, Citation Proximity Analysis
(CPA) maps the proximity between two citations in the full text to
the strength of the relevance of two co-cited papers. Two papers having
more strength of the co-cited relevance are more similar.

\item DeepWalk \cite{perozzi2014deepwalk}: DeepWalk is a learning algorithm
to obtain distributed representation for vertices in a network, which
was used in the original paper for relational classification problem.
As the paper citation relation is similar to a network while papers
can be regard as the vertices, DeepWalk can be easily introduced to
get paper vectors. The core idea in DeepWalk is to take random walk
paths from network as sentences, while the vertices as words. The
built corpus is then dropped to SkipGram to generate corresponding
distributed representation of vertices. DeepWalk declared a 10\% promotion
based on $F_{1}$ score on relational classification tasks than state-of-the-art
methods for social network representations. We transder the model to learn vectors of papers in citation network for our task.
\end{itemize}

\subsection{Evaluation}

With the calculated similarity scores by various algorithms, we can
get the rank of the most $K$ similar documents of every document
in the database. Because classical similarity methods cannot get the
similarity score for every pair arbitrarily, $K$ is not fixed, so
we conducted experiment under different $K$ to get a comprehensive
result. Intersection ratio are used as evaluation measurement for our experiment for
their invariance of $K$. Intersection ratio take the average ratio
of intersection between the top-$K$ document sets ranked according
to the similarity measure and the Mesh baseline respectively. 

\subsection{Results}

\begin{figure*}[tbp]
\begin{center}
\begin{raggedright} \subfloat[\label{fig:intersection1}PMCOS]{\begin{centering}
\includegraphics[width=6cm]{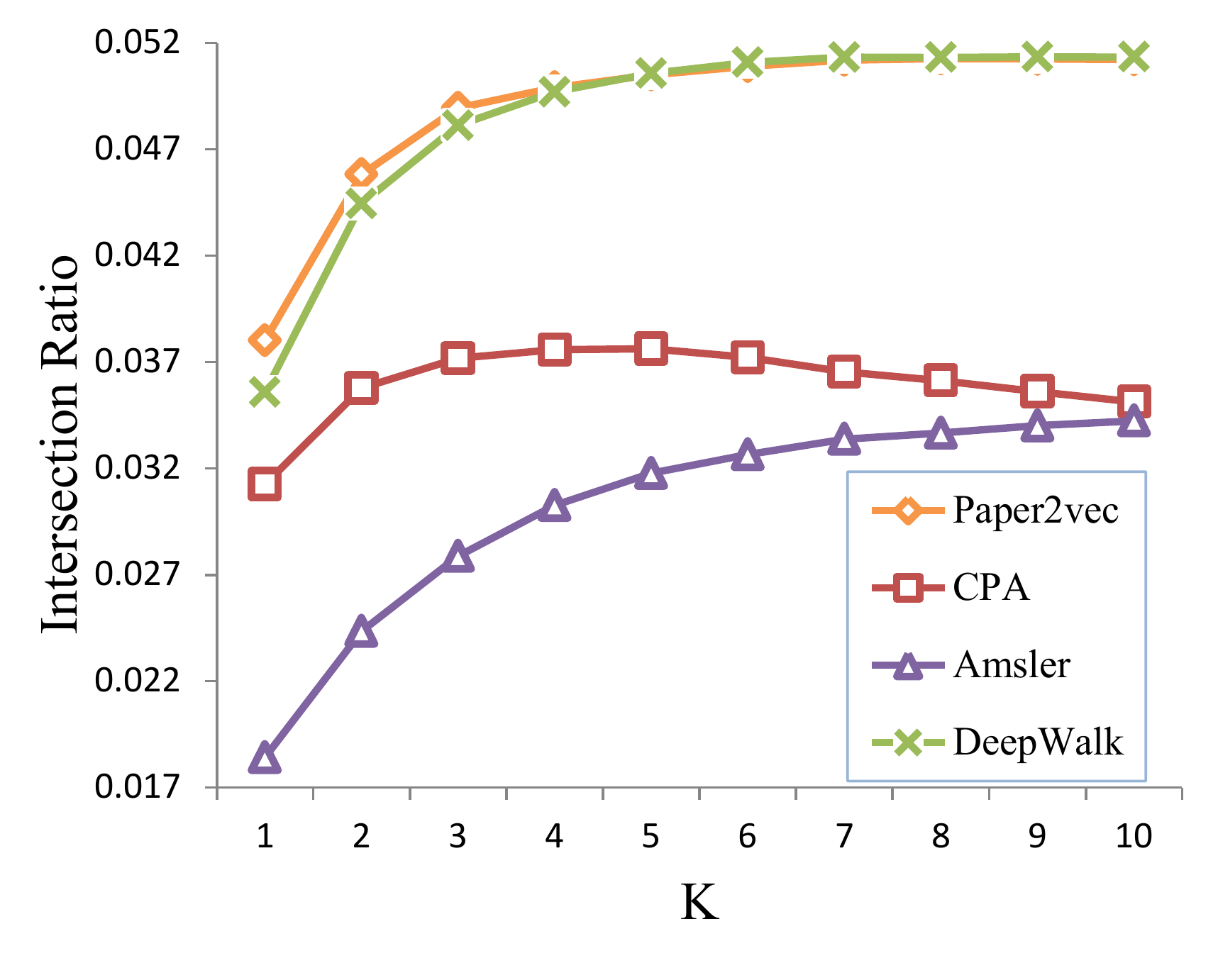}
\end{centering}
}
\end{raggedright} 
\begin{raggedleft} \subfloat[\label{fig:kendall1}TREC Genomics]{\begin{centering}
\includegraphics[width=6cm]{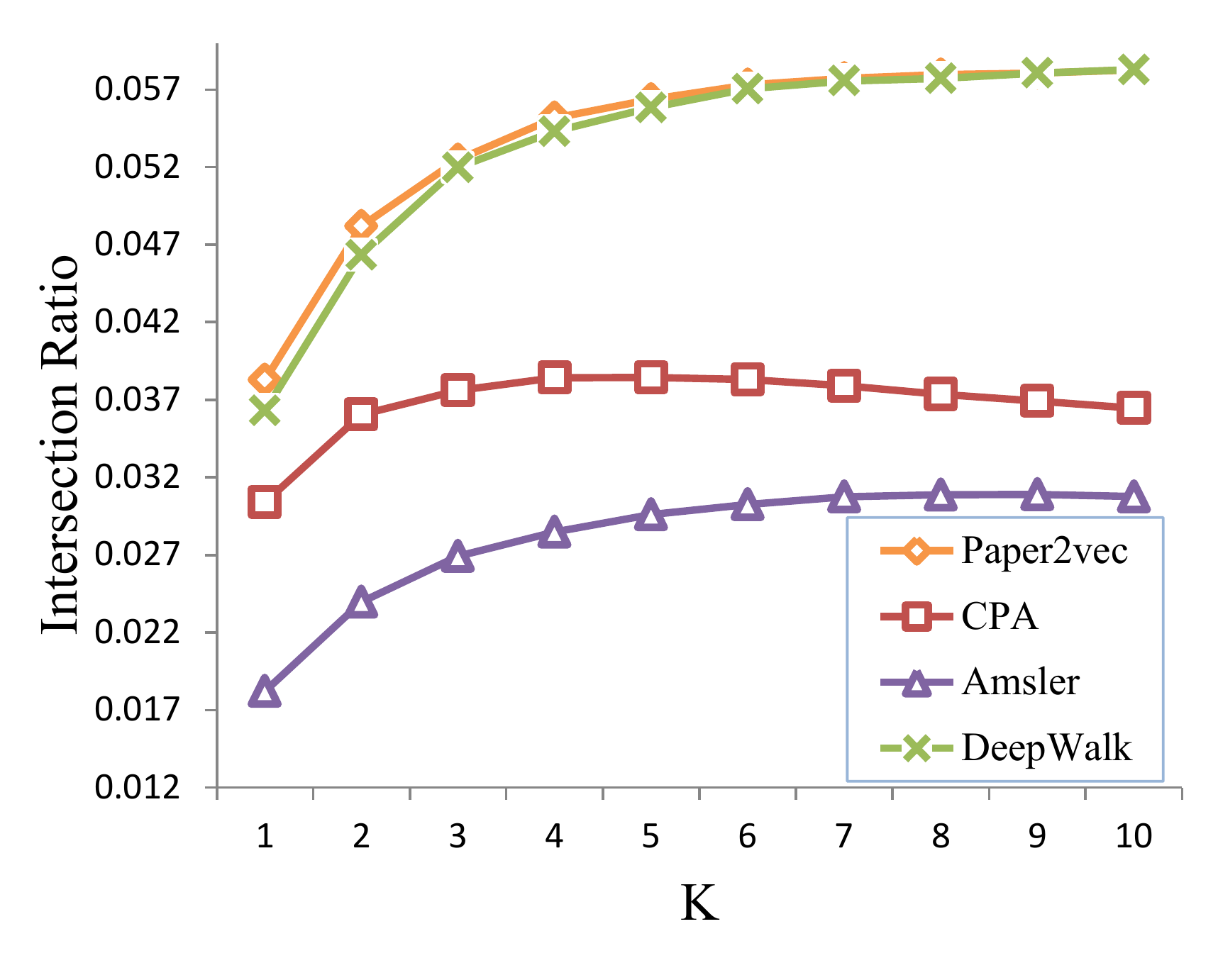}
\end{centering}
}
\end{raggedleft}
\end{center}
\caption{\label{fig:Evaluation-results-based}Evaluation results on both datasets on different $K$.}
\end{figure*}

We train 500-dimensional vectors for Paper2vec and DeepWalk and the
window size is both set 3. There are several variants of CPA model, we only list the best result around them.
The compared result under different $K$ is showed in Figure \ref{fig:Evaluation-results-based}.
DeepWalk and Paper2vec are both based on distributed representation and outperform other models significantly, which implies the promising
future of distributed representation in this area. Paper2vec is better than DeepWalk on small
$K$, meaning it can find better results in the first few documents, which is important for scholar recommendation. 

\subsection{Model Analysis: Window Size}

\begin{figure*}[tbp]
\begin{center}
\begin{raggedright} \subfloat[\label{fig:intersection}PMCOS]{\begin{centering}
\includegraphics[width=6cm]{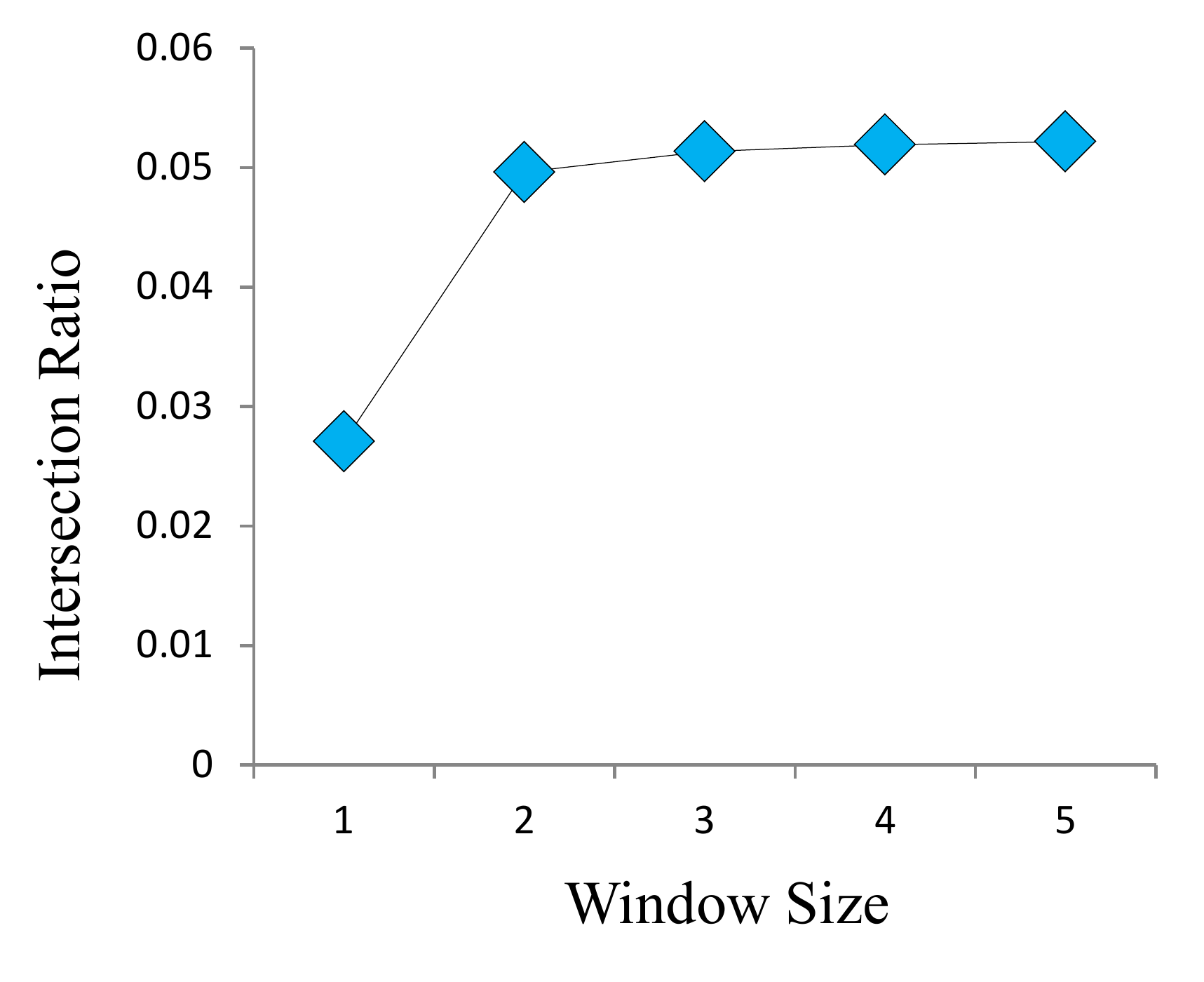}
\end{centering}
}
\end{raggedright} 
\begin{raggedleft} \subfloat[\label{fig:kendall}TREC Genomics]{\begin{centering}
\includegraphics[width=6cm]{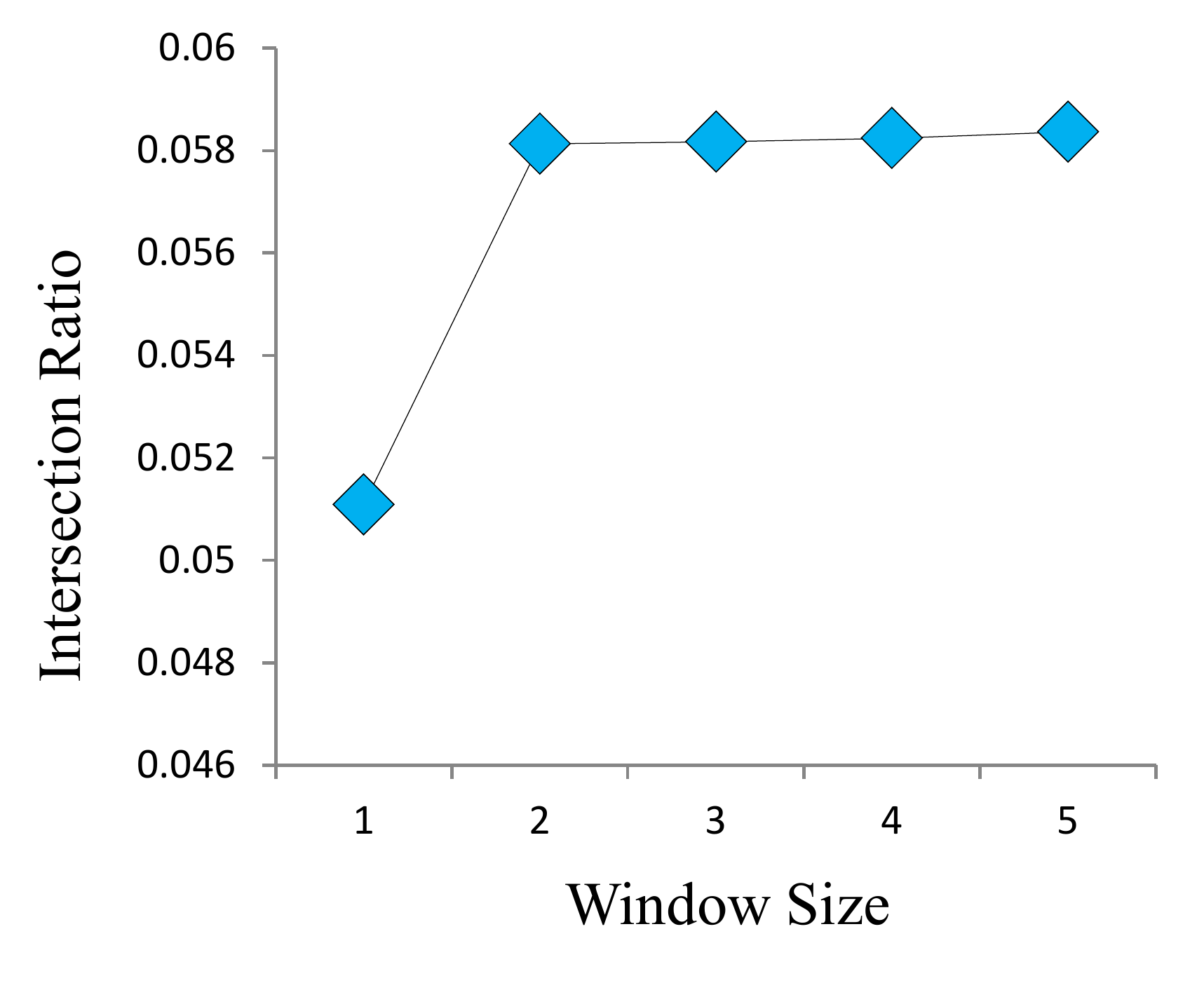}
\end{centering}
}
\end{raggedleft}
\end{center}
\caption{\label{fig:windowsize}Intersection ratio on datasets evaluation
as a function of window size of Paper2vec when $K=10$.}
\end{figure*}

Larger window should contains richer information about the context and results in better performance. So we look into the relation between the window size $win$ and the performance of the model. We trained Paper2vec model on datasets mentioned above for various training window size $win$. Parameters relating to training are the same as before and we consider the situation $K=10$. The result is showed in Figure \ref{fig:windowsize}. We can see a monotonic increase in performance as the window size $win$ increases, since larger context tends to contain more information about current document, as we supposed. With the increasing of window size, the marginal profit of information gained is diminishing and the curve slope is descending. The curve suggests the information distribution among the structure.

\subsection{Model Analysis: Novelty}
Novelty are highly desirable features for scholar recommendation, for the goal of scholar recommendation system is to help researchers find papers that are relevant but have not be found by themselves. However, models based on cooccurrences of links prefer items having more links. For example, CPA prefers popular papers that are cited frequently and all similar documents found by CPA should at least be cited at once. It is not the case for distributed representation based algorithms, which give every document a vector and just consider the distance between vectors. So we suppose our Paper2vec model tend to be more popularity independent than classical models. Inspired by novelty measurement in \cite{bellogin2010study}, we define a similar novelty measurement in a global perspective based on the concept of entropy in information theory. Considering the top-K similar documents found by all documents in the collection, given the collection set $S$, we get,
\begin{equation}
novelty=-\sum_{i\in{S}}{p_i\log{p_i}}
\end{equation}    
where,
\begin{equation}
p_i=\frac{{\vert \{j \vert i \in{R_j} \} \vert}}{\sum_k{\vert R_k\vert}}
\end{equation}    
$R_j$ denotes the top-K similar documents found by model for documnet j. The numerator part of the equation denotes the frequency document $i$ appears in other documents' similar lists. In information theory, the novelty measurement could be seen as the expect value of the information contained in the distribution of documents in the relevant set of all documents in the collection. The maximum value happens when all documents appear in relevant set at the same frequent, which is the ideal situation that there is no popular documents any more. More frequently one of the documents appears than others, smaller the novelty measurement, meaning model prefering some items than others when recommending. The measurement also decreases when the coverage of relavant set decreases. We name this measurement Entropy Novelty for it's derived from the concept of entropy in information theory. In equation $\vert R_k \vert$ is not fixed to K because in cooccurence model the size of similar documents set for every document is variant.

\begin{figure*}[tbp]
\begin{center}
\begin{raggedright} \subfloat[\label{fig:pmcos}PMCOS]{\begin{centering}
\includegraphics[width=6cm]{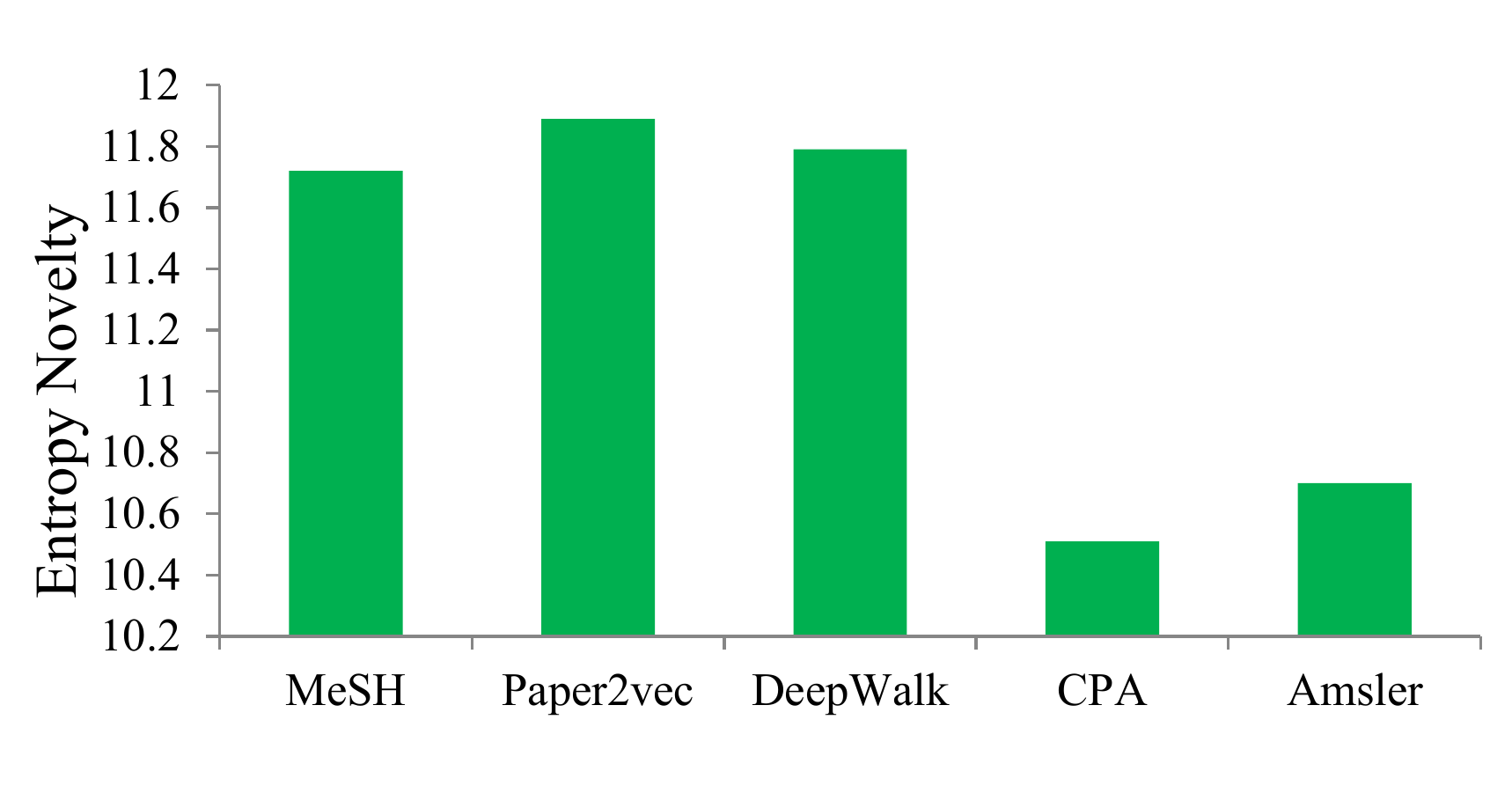}
\end{centering}
}
\end{raggedright} 
\begin{raggedleft} \subfloat[\label{fig:trec}TREC Genomics]{\begin{centering}
\includegraphics[width=6cm]{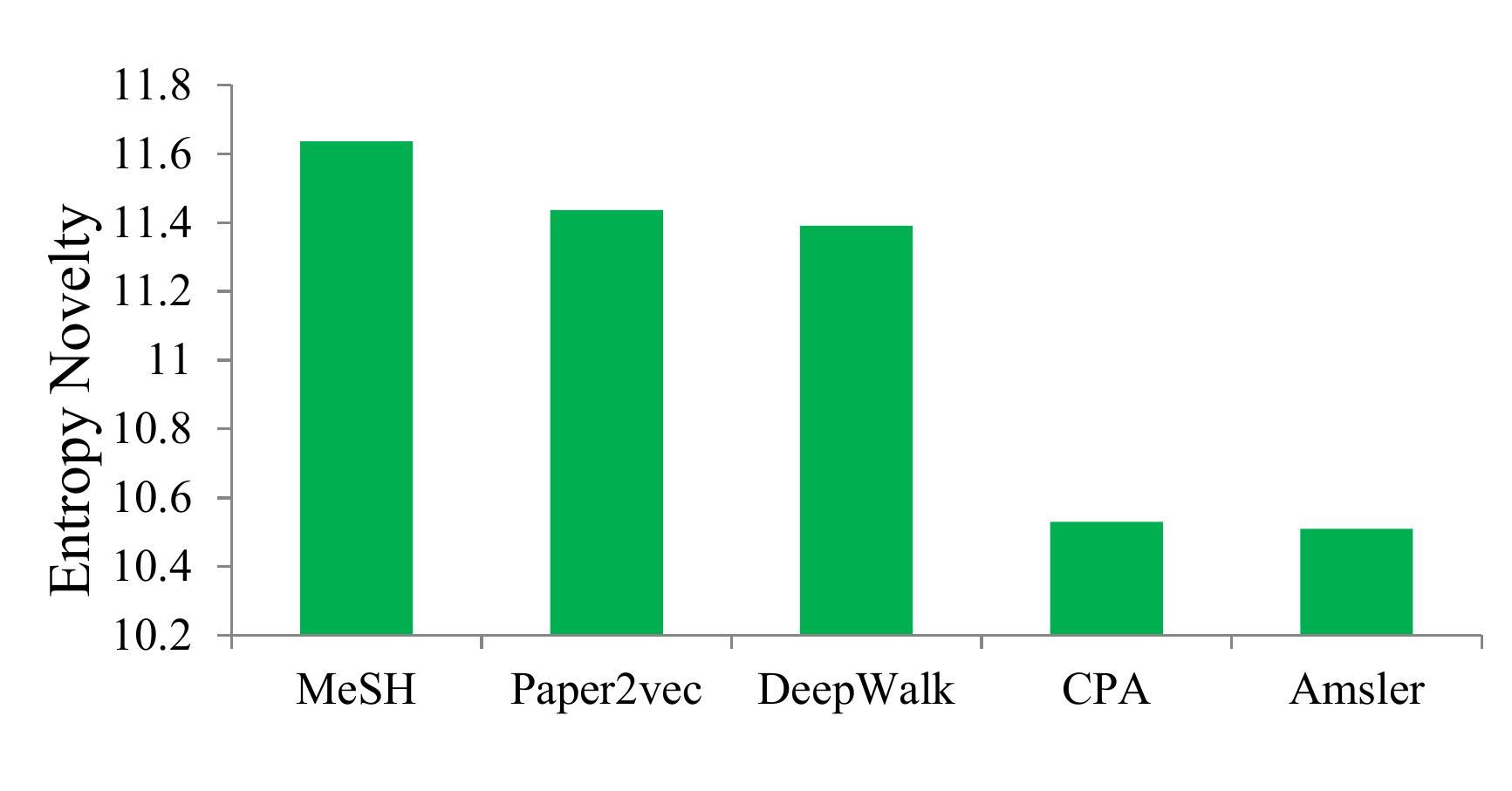}
\end{centering}
}
\end{raggedleft}
\end{center}
\caption{\label{fig:novelty}Entropy Novelty measurement on datasets.}
\end{figure*}

We calculated the Entropy Novelty for Paper2vec and other models mentioned in the baseline methods section on datasets PMCOS and TREC Genomics. The gold standard measurement MeSH is also calculated as control group. The result is showed on Figure \ref{fig:novelty}. While all distributed representation based models surpass cooccurence based models apparently, Paper2vec is the best around all models considered in both datasets, which proves distributed representation based models tend to consider semantic similarity of papers without the influence of other effects, such as popularity. This property can help users to find relevant papers that are hard to find by other classical models.

\section{Conclusion \& Discussion}

We proposed Paper2vec, a novel approach for learning latent distributed
representations from citation relations between documents reflecting
the topics. We define a weighted citation linkage context for papers
based on probability, and utilize a variant of matrix factorization to obtain
document distributed representation, which can be used for tasks such
as document clustering, relational classification, similarity measurement
and so on. For paper recommendation, the similarity measure of any
pair of documents can be calculated, the full text is not necessary,
and the stochastic training process make it possible to update the
new papers introduced into the database without training the whole
corpus again and easy to be parallelized. The advantages and better
performance of Paper2vec make it a promising method combined with
text-based method for future scholar recommender systems.

In addition, there are more untapped potential hidden in distributed
representation. \cite{Mikolov2013c} finds the vector difference of
distributed representation can suggest the similarity of pair of words.
For instance, vector(\textquotedblright{}King\textquotedblright{})
- vector(\textquotedblright{}Man\textquotedblright{}) + vector(\textquotedblright{}Woman\textquotedblright{})
results in a vector closed to vector(\textquotedbl{}Queen\textquotedbl{}).
\cite{levy2014linguistic} gave an math explanation of the property.
If the paper vectors have the same property, which can be used for
finding papers that having a specific topic relation with the input
document by simple vector algebraic operation. More research are needed
to verify this hypothesis.

\newpage{} \newpage{}

\bibliographystyle{plain}
\bibliography{all}

\end{document}